\documentclass[prd,aps,floats,preprintnumbers,preprint,nofootinbib,groupedaddress]{revtex4}
\usepackage{graphicx, epsfig}

\begin{document}
\begin{flushright}
MADPH-05-1422\\
HRI-P-05-04-001\\
hep-ph/0505260
\end{flushright}

\title{Neutrino masses and lepton-number violation\\ 
in the Littlest Higgs scenario}

\author{Tao Han}
\affiliation{Department of Physics,  University of Wisconsin, Madison,
Wisconsin 53706, USA}

\author{Heather E.\ Logan}
\thanks{Current address: Department of Physics, Carleton University,
1125 Colonel By Drive, Ottawa, Ontario K1S~5B6, CANADA}
\affiliation{Department of Physics,  University of Wisconsin, Madison,
Wisconsin 53706, USA}

\author{Biswarup Mukhopadhyaya}
\affiliation{Harish-Chandra
Research Institute,\\
Chhatnag Road, Jhusi, Allahabad - 211 019, INDIA}

\author{Raghavendra Srikanth}
\affiliation{Harish-Chandra
Research Institute,\\
Chhatnag Road, Jhusi, Allahabad - 211 019, INDIA}

\begin{abstract}
We investigate the sources of neutrino mass generation in Little Higgs
theories, by confining ourselves to the ``Littlest Higgs"  scenario. 
Our conclusion is that the most satisfactory way of incorporating neutrino
masses is to include a lepton-number violating interaction between the scalar 
triplet  and  lepton doublets. The tree-level neutrino 
masses generated by the vacuum expectation value of the triplet are found to
dominate over contributions from dimension-five operators so long as no
additional large lepton-number violating physics  exists at the
cut-off scale of the effective theory. We also calculate the various 
decay branching ratios of the charged and neutral scalar triplet  states, 
in regions of the parameter space consistent with the observed neutrino 
masses, hoping to search for signals of  lepton-number violating interactions
in collider experiments. 
\end{abstract}

\maketitle

\section{Introduction}
Little Higgs theories \cite{lhmodels,Nima,lhreview}
represent a new attempt to address the problem of quadratic divergence
in the mass of the Higgs boson responsible for 
electroweak symmetry breaking.  This approach treats the Higgs boson
as part of an assortment of pseudo-Goldstone bosons, 
arising from a global symmetry spontaneously broken at an energy 
scale $\Lambda$, typically on the order of 10 TeV. There is also an explicit 
breakdown of the overseeing global symmetry via gauge and 
Yukawa interactions, thereby endowing the Goldstone bosons with
a Coleman-Weinberg potential and making them massive.
The Higgs mass is thus protected by the global symmetries
of the theory and only arises radiatively due to the gauge and Yukawa 
interactions.
As an effective theory valid up to the scale $\Lambda$,
the model is rather economical in terms of the new fields introduced
in order to fulfill the necessary cancellation for the quadratic divergence
at the one-loop level.
The model requires, in addition to new gauge bosons and vectorlike fermions, 
the existence of additional scalars belonging to certain representations of the
Standard Model (SM) gauge group.

Aside from the crucial vector-like $T$-quark, the fermionic sector 
can essentially have the same appearance as in the SM.
There is no attempt made to address the origin of fermion masses 
and  mixing. In fact, the theory would encounter extremely stringent 
constraints from the absence of excessively large flavor-changing neutral
currents and CP violation in the fermionic sector \cite{flvr} if the scale 
responsible for flavor physics is at the order of the cutoff scale $\Lambda$. 
Flavor issues are thus ostensibly left out as problems awaiting the more
fundamental theory at higher energies, the so-called UV completion
of the theory, that would hopefully lead to the SM structure or similar
as an effective low-energy realization. 

However, one may like to remember that the only area 
where experimental hints of new physics have been found so far  
is the neutrino sector \cite{nu1}.
It is therefore both interesting and important to see if little Higgs 
theories can accommodate neutrino masses and mixing as suggested by 
the observed data.  It is even not unreasonable 
to say that it will be a vindication of little Higgs theories if they
at least suggest a mechanism  for the generation of neutrino masses. 
The present work aims to buttress this attempt.
Are the neutrinos acquiring their masses through interaction with new
particles already postulated in the theory?
What can be the detectable  signatures of the model 
carrying  imprints of the fact that its low-energy Lagrangian and
particle spectrum address the issue of neutrino masses?
We examine these questions by adopting the ``Littlest Higgs'' (LtH) 
model \cite{Nima}, which has been extensively studied in recent literature.

We explore the most economic extension of the basic model that is required to 
accommodate neutrino masses and is consistent with the demand that it does not 
affect the cancellation of  quadratic divergences in the SM Higgs mass. 
In particular, we make use of the fact that the LtH
scenario contains, in addition to the usual Higgs doublet, 
an additional set of scalars that form a complex triplet \cite{triple} 
under the SU(2) gauge group of the 
Standard Model with hypercharge $Y=1 \ (Q=I_3+Y).$
This complex triplet forms part of the assortment of Goldstone bosons when
a global SU(5) breaks down 
to SO(5) at the scale $\Lambda$ in this model.
There is an additional gauged SU(2)$\times$U(1) beyond that of the
SM, which is also spontaneously broken at scale $\Lambda$;
some of the aforementioned Goldstone bosons are absorbed as longitudinal
components of the extra gauge bosons. Ten scalar degrees of freedom
remain after this, and are found to consist of a doublet ($H$) and a complex 
triplet ($\phi$) under the electroweak SU(2). The complex triplet offers
a chance to introduce lepton number violating interactions into the theory.
We find that the most satisfactory way of incorporating neutrino
masses is to exploit such an interaction 
of the lepton doublets, leading to a Majorana mass for neutrinos and
lepton number violation by two units.
Then we proceed to examine the parameter 
range of this model consistent with the observed 
neutrino masses, and look at the 
consequence it has on the phenomenology of the model. 
In particular, we focus on the decays of the additional SU(2) triplet scalar 
states introduced in this scenario, which can 
have masses of order a TeV. We present calculations of the decay
branching ratios of the triplet states,  discuss the complementary roles
of different decay channels to test the scenario, and comment on 
their potential collider signatures within the region of the parameter 
space that is consistent with the observed neutrino masses.

Our paper is organized as follows.
In Section \ref{neumass}, 
the status of neutrino mass generation with a heavy right-handed
neutrino is first  briefly reviewed. We then 
take up the case of neutrino masses without any right-handed
neutrino, and show that the LtH construction
can accommodate the observed neutrino mass and mixing patterns. 
In particular, with the help of the complex triplet, 
one obtains dimension-4 lepton-number violating operators
($\Delta L = 2$).  The Majorana neutrino masses and their mixing 
can be generated by these operators consistent with current observations 
without necessarily pushing the couplings to tiny values; instead, the
smallness of the neutrino masses can be driven in part by a tiny
triplet vev. 
We also discuss the $\Delta L = 2$ operators with the full gauge
symmetry of the model and find that in such a scenario the couplings would 
have to be of order $10^{-11}$ to accommodate the observed neutrino masses.
In Section \ref{phis}, we 
study the decay channels of the triplets. 
These, we emphasize, constitute the characteristic signals 
of the triplet and allow a test of
the mechanism of neutrino mass generation.  We summarize and
conclude in Section \ref{four}. 
The features of the LtH scenario and the
interactions of the triplet that are relevant for our phenomenological
study are summarized in Appendix \ref{app:A}.  The triplet 
decay partial widths are listed in Appendix \ref{app:B}.

\section{Neutrino masses}
\label{neumass}

\subsection{Neutrino masses with right-handed neutrinos}

In the SM as well as the simplest little Higgs constructions, 
there are no right-handed
neutrino states that are singlets under SM gauge interactions.
By introducing right-handed neutrinos ($N_R$), one can obtain 
gauge-invariant Dirac mass terms from the SU(2) doublets of
the leptons $L$ and the Higgs $H$,
\begin{equation}
  y^D_{ij}\  \overline{L_{Li}}\  H^\dagger N_{Rj} + {\rm h.c.},
\end{equation}
with $i,j$ being generation indices,
as well as  Majorana mass terms
\begin{equation}
  - M_{ij} \overline{N_{Ri}^c} N_{Rj} + {\rm h.c.}
  = M_{ij} N_{Ri}^T C^{-1} N_{Rj} + {\rm h.c.},
\end{equation}
where $C$ is the charge-conjugation operator  
 in the notation of,  e.g., Ref.~\cite{Peskin}.

The Dirac terms alone  lead to a contribution to the
neutrino mass  of the order $m_\nu \sim y^D v$.  Since the neutrino
masses are known to be at most of order 0.3 eV \cite{pdg}, the
Yukawa couplings would have to be extremely small, $y^D \lesssim 10^{-12}$.
While technically natural, such tiny Yukawa couplings are difficult 
to rationalize.

Including the Majorana terms, light neutrino masses
are generated at the order $(y^D v)^2/M$ \cite{order}
by virtue of the well-known seesaw mechanism \cite{seesaw}.
If we assume that the Yukawa couplings $y^D_{ij}$ are naturally 
of the order of unity, then $M\gtrsim 10^{13}$ 
GeV in  order to obtain a neutrino mass less than about 0.3 eV.
The problem, however, is that if we take the Majorana mass scale
to be near the Little Higgs cutoff $\Lambda \simeq 10$ TeV,
then all of the Yukawa couplings would have to be quite small
and all roughly equal, $y^D_{ij}\lesssim 10^{-5}$ for all three generations.
This is in constrast to the corresponding charged leptons, for which
the Yukawa couplings exhibit a large hierarchy
between generations. Of course, the right-handed neutrino mass that determines
the seesaw scale could be much higher than $\Lambda$, as in the usual
seesaw scenario within the Standard Model.
However, in this work we wish to look for alternative explanations of the
neutrino masses within the context of the LtH scenario with 
observable signatures that 
do not rely upon physics above the cutoff scale $\Lambda$.

\subsection{Neutrino masses in the absence of right-handed neutrinos}
\label{sec:lphil}

To us, the solution seems to be in avoiding the introduction of massive
right-handed neutrinos altogether in a little Higgs scenario. One can
still construct Majorana mass terms with the help of the Higgs triplet
in the LtH model, obtained from a dimension-four
$\Delta L = 2$  coupling,
\begin{equation}
	{\cal L} = iY_{ij} L_i^T \ \phi \, C^{-1} L_j + {\rm h.c.}
	 \label{lphil}
\end{equation}
Note that the definition of $\phi$ here includes  $(-i)$, as evident from
Eq.~(\ref{eq:Hphixi}).
With the vacuum expectation value (vev) of $\phi^{0}$ being $v^{\prime}$, 
the induced neutrino masses are of 
the order of $Yv^{\prime}$. With a sufficiently small $v^{\prime}$, 
as preferred by the precision electroweak data \cite{EWvprime}, 
adequate neutrino masses may be generated.  The occurrence
of such Majorana masses has already been discussed in the context of
general models with triplet scalars \cite{triple,trip}.  

In the LtH model, however, some additional caution is necessary, since here 
we have an effective theory with a rather low cut-off.
It can be argued that, if there is lepton-number violating physics at the scale
$\Lambda$, then it is practically impossible to prevent the appearance of
dimension-five operators of the form
\begin{equation}
	Y_5 { (HL)^2\over \Lambda }
	\label{eq:dim5op}
\end{equation}
giving rise to neutrino masses on the order of $Y_5 v^2/\Lambda$.
This contribution to the neutrino masses is inadmissibly large if $Y_5$ is 
naturally of order unity.
Of course, one may suppress the neutrino
mass by requiring that the seesaw scale 
corresponding to lepton-number violation
is not $\Lambda \ (\sim 10$ TeV) but some higher 
scale, perhaps corresponding to a
grand unification scenario. However, as we have mentioned above, this
solution is somewhat unsatisfying in the little Higgs context,
since the entire issue of grand unification is unclear in a
UV-incomplete theory.

The way out of the difficulty is to postulate 
that there is {\em no additional lepton-number violating physics
at the scale $\Lambda$}, and that the {\em only} $\Delta L = 2$ 
effect comes from the coupling given by Eq.~(\ref{lphil}).
Such a postulate is plausible in the sense that the operator of  
Eq.~(\ref{lphil}) is renormalizable and independent of the cutoff.
Such a postulate also keeps the scenario
minimal in terms of particle content, since right-handed neutrinos, unlike
the scalar triplets, do not arise from any intrinsic requirement 
of the model. 
The absence
of right-handed neutrinos at or below the scale $\Lambda$ 
prevents the potentially dominant dimension-five
operators of Eq.~(\ref{eq:dim5op}). Such operators can then arise only through 
loop effects involving the $\Delta L = 2$
couplings of the $\nu_L$ to the scalar triplet.  
As we demonstrate below in Sec.~\ref{sec:constraints}, the structure
of the Coleman-Weinberg potential ensures that the contributions of these
loop-induced dimension-five operators to the neutrino masses are subleading 
compared to the tree-level $\Delta L = 2$ interaction given above.

Thus, neutrino masses are perhaps best implemented in the LtH model
solely in terms of the tree-level $\Delta L = 2$ interaction of the scalar 
triplet.  So far there is no need to attribute the effect to
a high scale, since lepton-number conservation is not dictated by
any underlying symmetry of the theory. 
The relevance of this term is further accentuated by the fact
that the triplet vev in any case has to be quite small 
compared to the electroweak scale, in order to be consistent with the
limits on the $\rho$-parameter \cite{EWvprime,rho}. 
Thus, seeds of small neutrino masses can already be linked to the
electroweak precision constraints.

It should be noted that although the $LL\phi$ interaction term is invariant
under the standard SU(2)$_L \times$U(1)$_Y$ symmetry, it breaks the full
[SU(2)$\times$U(1)]$^2$ gauge invariance of the LtH model.  The $LL\phi$ 
interaction term is invariant under the two U(1) symmetries so long as
the U(1) charges of the lepton doublet are chosen to cancel anomalies in the
full theory (see Sec.~\ref{sec:largersym} for details).  
On the other hand, this term 
breaks the [SU(2)]$^2$ part of the full gauge symmetry because the triplet
$\phi$ is a Goldstone boson of the full theory and transforms nonlinearly
under the two SU(2)s, while $L$ transforms as a doublet under only SU(2)$_1$.
We note however that the real motivation for this enlarged gauge
symmetry is the cancellation of potentially large quadratically divergent
contributions to the Higgs mass.  Apart from that, there is no requirement
that such an invariance holds in all sectors of the theory.

It can be seen through explicit calculation that the cancellation 
of quadratic divergences is {\em not} affected so 
long as the non-invariance under
[SU(2)$ \times $U(1)]$^2$ is confined only to the lepton-number
violating interaction of the triplet.  In particular, the global symmetry
structure in the gauge and top-quark sectors that protects the Higgs 
mass at one-loop level is not affected by the new $LL\phi$ interaction.
The only effect of this interaction on the Coleman-Weinberg potential
\cite{cw} for the scalars
is a contribution to the coefficient of the triplet mass, $\lambda_{\phi^2}$
(see Appendix \ref{app:A} for details), for which the  modified one-loop 
expression is 
\begin{equation}
	\lambda _{\phi ^2} = \frac {a}{2}\left[ \frac {g^2}{s^2c^2}+
	\frac {g^{\prime 2}}{s^{\prime 2}c^{\prime 2}} \right]
	+ 8 a^{\prime} \lambda_1^2 + a^{\prime\prime} {\rm Tr}(YY^\dag ),
	\label{eq:lambdaphi2}
\end{equation}
where $a^{\prime\prime}$ is an arbitrary $\mathcal{O}(1)$ constant reflecting
the UV incomplete nature of the theory.  The overall constraint to be satisfied
by the modified expression is that $\lambda_{\phi^2}$ should remain positive, 
so that the triplet vev, purportedly small, is generated through 
doublet-triplet mixing only.  If $a^{\prime\prime}$ is positive, it
results in a slight enhancement of the triplet scalar mass compared
to that in the minimal LtH scenario. Thus the introduction of the
$LL\phi$ interaction seems to be consistent with the fundamental spirit of
the little Higgs approach. We lay out the full interaction terms of
Eq.~(\ref{lphil}) in Appendix \ref{app:A} for future phenomenological 
considerations.

\subsection{Constraints from neutrino masses}
\label{sec:constraints}

Since our first concern is to see the viability of this proposal, we 
begin by assuming neutrino masses to be of order 0.1 eV.
The left-handed Majorana neutrino mass matrix resulting from
Eq.~(\ref{lphil}) in this scenario is
\begin{equation}
	{\cal M}_{ij} ~=~Y_{ij}v^{\prime} .
\end{equation}
We neglect CP-violating phases.  Then $Y$ is a ($3\times 3$) symmetric 
matrix with six independent parameters.
The physical neutrino masses are the product of $v^{\prime}$ and
the eigenvalues of $Y$. The elements of $Y$ can in principle be as 
large as perturbation theory permits; we consider them to have a natural 
size of order unity.  The triplet vev $v^{\prime}$ is restricted 
to be $\lesssim 1$ GeV from the constraints on the 
$\rho$-parameter \cite{EWvprime,rho}. 

The smallness of  the neutrino masses leaves us with two extreme
alternatives, as described below.
\begin{enumerate}
	\item The elements of $Y$ are very small, typically of the 
	order $10^{-10}$, and $v^{\prime} \sim 1$ GeV. This means
        that the $LL\phi$ interaction term in Eq.~(\ref{lphil})
        supplies the physics responsible for the smallness of neutrino masses.
	\item $Y\simeq 1$ together with an extremely small $v^{\prime}$, 
	arising from a tiny value of the induced doublet-triplet mixing 
	coefficient $\lambda_{h \phi h}$ in the Coleman-Weinberg potential.
        In this case the Coleman-Weinberg potential provides the physics
        behind the smallness of neutruino masses, while the origin of
        bi-large mixing has to be sought in the relative values of the
        different $Y_{ij}$. 
\end{enumerate}
The first option leads to very small couplings, which could be argued to
be unnatural. One needs to remember that the physics linked with
$Y_{ij}$ is {\em not only lepton-number violation but also lepton-flavor 
violation}. Therefore the coupling in Eq.~(\ref{lphil}) must
have its origin at a scale much higher than $\Lambda$, in order to 
avoid unacceptable flavor
violation in the low-energy theory and the appearance of large dimension-five
operators.  Thus the explanation for the smallness of the neutrino masses
is pushed up to scales much higher than $\Lambda$.

The second scenario, on the other hand, has a certain 
advantage.  In addition to generating neutrino masses of the right order, 
one also has to explain the observed bi-large mixing pattern in the 
neutrino sector. A model-independent fit of such mixing requires one to 
fine-tune the elements of $Y$.  Having all six elements on the order 
of $10^{-10}$ enhances the degree of fine-tuning even further. It may 
therefore be a slightly less disquieting prospect 
to envision the ``fine-tuned'' elements of $Y$ as being close to unity, and 
have a very small vev for the triplet.  The generation of such a small
vev must be accomplished by appropriate values of the parameters that
determine $\lambda_{h \phi h}$ at the scale $\Lambda$.  As can be seen from
the detailed expressions listed in Appendix~\ref{app:A1}, 
a small triplet vev can arise,
for example, from a cancellation of the gauge and Yukawa contributions
to the Coleman-Weinberg potential. While a theoretical explanation has to 
await the UV completion of the scenario, this situation is consistent with 
all other aspects of the model, and has distinctive phenomenological 
implications. Thus we have chosen to explore such implications in detail,
remembering all along that the final explanation for the smallness of 
the neutrino masses is linked to the UV completion of the LtH model.

To summarize, we will concentrate only on the operator of 
Eq.~(\ref{lphil}), with the requirement 
$m_{\nu}\simeq Yv^{\prime}\simeq 10^{-10}$ GeV. Within this constraint, 
the $\Delta L~=~2$ coupling $Y$ [which is actually a ($3 \times 3$) matrix] 
and the triplet vev $v^{\prime}$ 
can vary over a wide range in our formulation. 
As we shall see in the next section, the phenomenological consequences
are especially interesting in the parameter ranges  
\begin{equation}
	10^{-5} < Y_{ij} \lesssim 1,\qquad \qquad 
	0.1\ {\rm MeV} > v^{\prime} > 1\ {\rm eV}.
	\label{range}
\end{equation}

\begin{figure}
\includegraphics{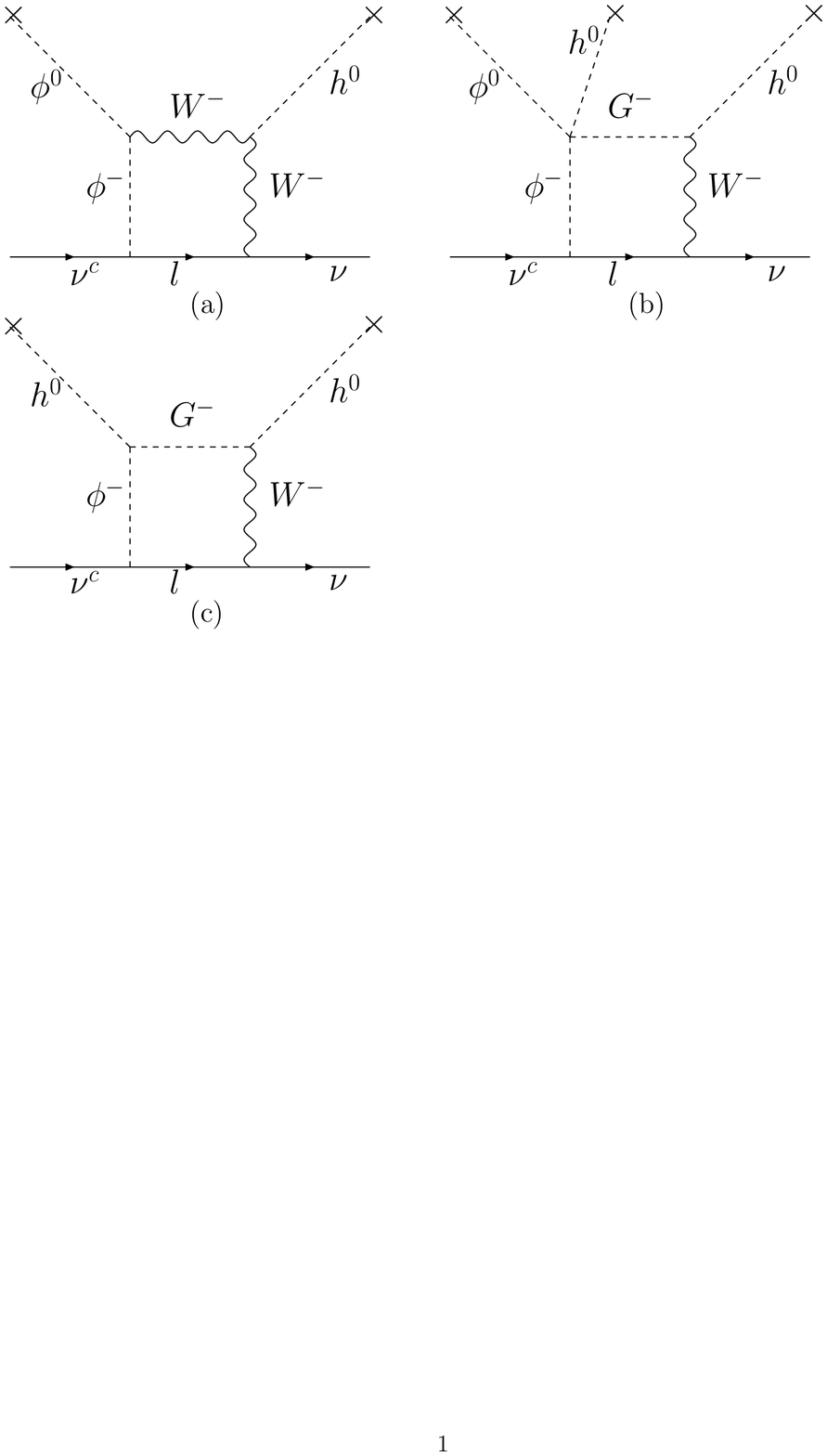}
\caption{ Representative one-loop diagrams giving rise to neutrino masses 
via dimension-five operators. }
\label{fig:one}
\end{figure}

It is important not to overlook other potentially significant  contributions 
to the neutrino masses through dimension-five operators induced at the 
one-loop level. Some representative diagrams leading to such operators are 
shown in Fig.~\ref{fig:one}, 
where we have worked in the 't~Hooft-Feynman gauge.  All of these diagrams 
give neutrino masses of the form $M_{ij} \nu_{Li}^T C^{-1} \nu_{Lj}$.

The neutrino mass from Fig.~\ref{fig:one}(a) is
\begin{equation}
	\mathcal{M}_{ij} = i v v^{\prime}
	\frac{M_Wg^3Y_{ij}}{\sqrt{2}}
	\int \frac{d^4p}{(2\pi )^4}\frac{1}{(p^2-m^2_{\phi })(p^2-M^2_W)^2}
        \approx 
	(Y_{ij}v^{\prime})\frac{g^4\ v^2}{32\sqrt{2}\pi ^2m^2_{\phi }}.
\end{equation}
Clearly this is a subleading contribution compared
to $Y_{ij} v^{\prime}$, being suppressed by a loop
factor times $v^2/m_{\phi}^2$. Similarly, the contribution from 
Fig.~\ref{fig:one}(b) is
\begin{equation}
	\mathcal{M}_{ij}\approx 
	- (Y_{ij}v^{\prime})\frac{g^2\lambda _{h\phi \phi h}v^2}
	{32\sqrt{2}\pi ^2m^2_{\phi }}
	= (Y_{ij}v^{\prime}) \frac{g^2 v^2}{24 \sqrt{2} \pi^2 f^2},
\end{equation}
where we have used the relation 
$\lambda_{h\phi \phi h} = -4 \lambda_{\phi^2}/3 = -4 m_{\phi}^2 / 3 f^2$
(see Appendix \ref{app:A} for details).
This is again suppressed by a loop factor times $v^2/f^2$.

The contribution from Fig.~\ref{fig:one}(c) is 
\begin{equation}
	\mathcal{M}_{ij} = i\frac{g^2}{4}\frac{Y_{ij}}{\sqrt{2}}
	\lambda_{h\phi h}fv^2 \int \frac{d^4p}{(2\pi )^4}
	\frac{1}{(p^2-m^2_{\phi })(p^2-M^2_W)^2}
	\approx  (Y_{ij} v^{\prime}) \frac{g^2}{32 \sqrt{2} \pi^2},
\end{equation}
where we have made use of the relation
$v^{\prime} = \lambda_{h\phi h}v^2/2\lambda_{\phi ^2}f$
from the minimization conditions of the Coleman-Weinberg potential
and $\lambda_{\phi^2} f^2 \simeq m_{\phi}^2$ (see Appendix \ref{app:A}).
Unlike the other diagrams, the neutrino mass contribution from this diagram 
is suppressed by only a loop factor.  This is bacause the size of the
diagram in Fig.~\ref{fig:one}(c) is controlled by the doublet vev $v$.  
The special relationships among parameters in the Coleman-Weinberg potential 
enables one to re-express the contribution in terms of the triplet 
vev $v^{\prime}$.
The contribution is larger than that from Figs.~\ref{fig:one}(a) and (b), 
although the loop factor ensures that it is subleading. Such a contribution
can play a potentially important role in determining the precise values of 
the neutrino mixing angles.

There is also a diagram in which the vertical $W^-$ propagator 
in Fig.~\ref{fig:one}(a) is replaced 
by the charged Goldstone $G^{-}$.  Since the 
coupling of $G^{-}$ to $\bar{\nu },\ell$ is suppressed by $m_{\ell}/M_W$, this 
diagram gives only a small contribution.  Similarly, diagrams involving
a virtual $Z$ boson and neutrinos are negligible.
Finally, the above expressions are subject to additional corrections due to
doublet-triplet mixing, which are further suppressed by $v^{\prime}/v$.

\subsection{${\mathbf \Delta L = 2}$ operators with larger symmetry}
\label{sec:largersym}

Thus far, our approach to constructing lepton-number violating operators
has been guided by the SM gauge invariance and naturalness considerations,
subject to the experimental constraints on the neutrino masses and mixing
patterns. The treatment of the scalar triplet separate from the doublet 
requires some mechanism to split the interactions of these two components
of the non-linear $\Sigma$ field, which is beyond the scope
of our phenomenological considerations in the current work. Nevertheless,
it is tempting to ask if one can instead construct operators that respect the
full gauge symmetry of the LtH model, namely 
[SU(2)$\times$U(1)]$^2$ gauge invariance.

Following the conventions of Refs.~\cite{Nima,han},
in which the third-generation quark doublet is extended to
$\chi^T = (b_L \ t_L \ T_L)$,
we write the lepton doublets as $L^T = (\ell_L \ \nu_L)$.
We can then write down the following lepton flavor violating operator,
\begin{equation}
  \mathcal{L}_{LFV} = -\frac{1}{2} Y_{ij} f 
  \left( L_i^T \right)_{\alpha}
  \Sigma^*_{\alpha\beta} C^{-1} \left( L_j \right)_{\beta} + {\rm h.c.},
  \label{eq:LFV}
\end{equation}
where $i,j$ are generation indices and $\alpha,\beta=1,2$ are SU(5) indices.
This operator is gauge invariant under both the SU(2)$_{1,2}$ gauge groups and 
under hypercharge.
This operator is also gauge invariant under both of the U(1)$_{1,2}$
gauge groups if the lepton charges under the two U(1) groups are
given by $Y_1(L) = -3/10$ and $Y_2(L) = -1/5$.  In the notation of
Ref.~\cite{han}, this corresponds to $y_e = 3/5$, as 
shown in Table~\ref{Tab:U1}.
\begin{table}
\begin{center}
\begin{tabular}{lccc}
\hline \hline
 & $\Sigma^*_{\alpha\beta}$ & $L$ 
	& $L^T_{\alpha} \Sigma^*_{\alpha\beta} C^{-1} L_{\beta}$ \\
\hline
U(1)$_1$ & $3/5$ & $3/10 - y_e$ & $6/5 - 2 y_e$ \\
U(1)$_2$ & $2/5$ & $-4/5 + y_e$ & $-6/5 + 2 y_e$ \\
Hypercharge & $1$ & $-1/2$ & $0$ \\
\hline \hline
\end{tabular}
\end{center}
\caption{Charge assignments of the lepton and scalar fields
and of the operator in Eq.~(\ref{eq:LFV}) under
the two U(1) gauge groups and hypercharge, with $\alpha,\beta = 1,2$.}
\label{Tab:U1}
\end{table}
This is the same condition that ensures anomaly cancellation
among the SM fermions.  This can be understood as follows.  The anomaly
cancellation condition is satisfied when the U(1)$_{1,2}$ charges of 
the fermions
are proportional to their hypercharges.  Since the operator in 
Eq.~(\ref{eq:LFV}) conserves hypercharge, the anomaly-free condition is 
sufficient to ensure that this operator also conserves the U(1)$_{1,2}$ 
charges.

Expanding the upper two-by-two block of the matrix $\Sigma^*_{\alpha\beta}$
in terms of the scalar fields $H$ and $\phi$ (see Appendix A), 
we have
\begin{equation}
  \Sigma^*_{\alpha\beta} = -\frac{2}{f} \left( \begin{array}{cc}
    \phi^{++} & \phi^+/\sqrt{2} \\
    \phi^+/\sqrt{2} & \phi^0 \end{array} \right)
  - \frac{1}{f^2} \left( \begin{array}{cc}
    h^+ h^+ & h^+ h^0 \\
    h^+ h^0 & h^0 h^0 \end{array} \right)
  + \cdots
\end{equation}
Inserting this into Eq.~(\ref{eq:LFV}), we obtain
\begin{eqnarray}
  \mathcal{L}_{LFV} &=& Y_{ij} \left[
    \nu_{Li}^T C^{-1} \nu_{Lj} \left( \phi^0 + \frac{1}{2f} h^0 h^0 \right)
    + \left( \nu_{Li}^T C^{-1} \ell_{Lj} + \ell_{Li}^T C^{-1} \nu_{Lj} \right)
    \left( \frac{1}{\sqrt{2}} \phi^+ + \frac{1}{2f} h^+ h^0 \right)
    \right. \nonumber \\ && \left.
    + \ell_{Li}^T C^{-1} \ell_{Lj}
    \left( \phi^{++} + \frac{1}{2f} h^+ h^+ \right) \right]
  + {\rm h.c.}
\label{eq:LFVlagrangian}
\end{eqnarray}
Clearly, the nonlinear sigma model has served to relate the dimension-four
$\overline{\nu^c_i} \nu_j \phi^0$ coupling to the dimension-five
$\overline{\nu^c_i} \nu_j h^0 h^0$ coupling.
This gives rise to a mass matrix for the neutrinos involving both 
$v^{\prime}$ and $v$:
\begin{equation}
	\mathcal{M}_{ij}
	= Y_{ij}\left( v^{\prime}+\frac{v^2}{4f}\right).
\end{equation}
Equation (\ref{eq:LFVlagrangian}) gives, to the leading order, the  
following dimension-four couplings of scalars to left-handed lepton
pairs:
\begin{eqnarray}
  \mathcal{L}_{LFV}^{dim-4} &=& Y_{ij} \left\{
    \ell_{Li}^T C^{-1} \ell_{Lj} \phi^{++}
    + \frac{1}{\sqrt{2}} 
    \left( \nu_{Li}^T C^{-1} \ell_{Lj} 
	+ \ell_{Li}^T C^{-1} \nu_{Lj} \right) \phi^+
    + \nu_{Li}^T C^{-1} \nu_{Lj} \phi^0
	\right\} \nonumber \\
	&+& Y_{ij}^* \left\{
	\overline{\ell_{Li}} \, C \, \overline{\ell_{Lj}}^T \phi^{--}
	+ \frac{1}{\sqrt{2}}
	\left( \overline{\nu_{Li}} \, C \, \overline{\ell_{Lj}}^T 
	+ \overline{\ell_{Li}} \, C \, \overline{\nu_{Lj}}^T \right) \phi^-
	+ \overline{\nu_{Li}} \, C \, \overline{\nu_{Lj}}^T \phi^{0*} \right\},
	\label{app:lphil}
\end{eqnarray}
where in the second line we have explicitly written out the Hermitian
conjugate piece.
Note that $\phi^0$ is a complex field containing real scalar and pseudoscalar 
degrees of freedom, $\phi^0 = (\phi^s + i \phi^p)/\sqrt{2}$.

The expression (\ref{eq:LFV}) is invariant under the full
[SU(2)$\times$U(1)]$^2$ gauge symmetry and preserves the 
nonlinear sigma model form for the scalar interactions.
However, the price to pay in such an approach is that one has to 
include dimension-5 terms proportional to $H^2$ from the beginning,
and thus have contributions to the neutrino masses proportional to $v^2/f$. 
Unlike the dimension-5 operators generated by the diagrams
in Fig.~\ref{fig:one}, these contributions are not 
proportional to $Y_{ij} v^{\prime}$ times a loop
suppression factor and cannot in general be made small, since
$f\simeq$ TeV if we have to stabilize the Higgs mass. As a result, 
this approach almost invariably ends up requiring values
\begin{equation}
	Y_{ij}\sim 10^{-11},
	\label{eq:z}
\end{equation}
for the $\Delta L = 2$ couplings of all $i, j$.
They are indeed unnaturally small. This implies the need for a more
fundamental explanation for neutrino masses beyond the effective theory
at the scale $\Lambda$. 

On the other hand, 
in our approach of separating the lepton-number violating couplings of 
$\phi$ and $H$, one can avoid extreme fine-tuning of 
the $Y_{ij}$ couplings and at the same time ensure neutrino masses 
of a size consistent with experimental data.
This is because our starting point is the 
dimension-four renormalizable operator of
Eq.~(\ref{lphil}), as opposed to the higher dimensional ones discussed in 
the alternative approach.  Thus our formulation by keeping only the
L-violating terms of Eq.~(\ref{lphil}) is independent of the cut-off. 
It is admittedly a phenomenological approach, and assumes that, whatever be 
the mechanism responsible for the breakdown of [SU(2)$\times$U(1)]$^2$
in the L-violating sector, any additional induced term proportional to
$(v^2/f)$ is suppressed.  
We nonetheless feel that this approach is quite general 
and model-independent, especially because the cancellation of quadratically 
divergent contributions to the SM Higgs mass remains unaffected, as was
discussed in Sec.~\ref{sec:lphil}.

We take this opportunity to note that an attempt has been recently made
in Refs.~\cite{kilian,nuh1} similar to the approach of Eq.~(\ref{eq:LFV}).
In Ref.~\cite{nuh1}, this operator was given in the form
\begin{equation}
	{\cal L}_{LFV} = z_{ij} \epsilon^{\alpha\beta}\epsilon^{\gamma\delta} 
	f \left( L_i^T \right)_{\alpha} \Sigma^*_{\beta\gamma}
	C^{-1} \left( L_j^T \right)_{\delta}  
	+ {\rm h.c.},
	\label{eq:Lee}
\end{equation}
which is equivalent to our result if $L^T = (-\nu \ \ell)$ is
used in Eq.~(\ref{eq:Lee}).
The authors of  Refs.~\cite{kilian,nuh1} also found the same conclusion as in
Eq.~(\ref{eq:z}), that $Y_{ij} \sim 10^{-11}$.

\section{Decays of the triplet states}
\label{phis}

We now examine the observable consequences of the scalar 
triplet having a vev and lepton number violating interactions compatible with
the observed neutrino masses.  In particular, we consider the decays of the 
scalar triplet into various characteristic final states, and discuss
their observable signals in future collider experiments. 

First of all, we note that the mechanism of scalar mass generation through 
the Coleman-Weinberg mechanism \cite{cw}
in the LtH model implies that 
the members of the triplet, $\phi^{++}$, $\phi^+$, $\phi^s$, and $\phi^p$
(where $\phi^s$ and $\phi^p$ are the scalar and pseudoscalar components
of $\phi^0$), are 
degenerate at lowest order with a common mass $m_{\phi}$.  
Their masses are split by electroweak symmetry breaking effects, 
leading to masses $m_{\phi}[1 + \mathcal{O}(v^2/m_{\phi}^2)]$.
The mass splittings are thus quite small for $m_{\phi} \gg M_W$, 
and we will neglect them 
in what follows.  The relevant interaction terms for the 
$\Delta L=2$ processes are given in Table~\ref{tab:feynmanDL2} in 
Appendix \ref{app:A}. 
The other $\phi$ couplings conserving the lepton number have been given
in Ref.~\cite{han}.  For completeness, they
are also tabulated in Table~\ref{tab:feynman} in Appendix \ref{app:A}.  
The possible decays of the triplet states are
\begin{eqnarray}
  \phi^{++} &\to& \ell^+_{i} \ell^+_{j},\quad  W^+ W^+,
	\nonumber \\
  \phi^+ &\to& \ell^+_{i} \bar{\nu}_{\ell_{j}},\quad t\bar{b},\quad  
T\bar{b},\quad W^+ Z, \quad W^+ h,
	\nonumber \\
  \phi^s &\to & \nu_{i} \nu_{j}, \quad \bar \nu_i \bar \nu_j, 
  \quad  t\bar{t},\quad  b\bar{b},\quad
 t{\bar T}+{\bar t}T,\quad  ZZ, \quad hh,
	\nonumber \\
  \phi^p &\to & \nu_{i} \nu_{j}, \quad \bar \nu_i \bar \nu_j, 
  \quad  t\bar{t},\quad  b\bar{b},\quad
 t{\bar T}+{\bar t}T, \quad Zh.
\end{eqnarray}
The full set of partial decay widths is listed in Appendix \ref{app:B}.  

To clearly see the interesting physics points, we discuss 
the partial decay widths for the doubly-charged Higgs boson
for $m_\phi \gg M_W$,
\begin{eqnarray}
  \Gamma (\phi^{++}\rightarrow \ell^+_i \ell^+_i) 
  = \frac{|Y_{ii}|^2 m_\phi}{8\pi }  ,\  \
  \Gamma (\phi^{++}\rightarrow W^+_L W^+_L) \approx
  \frac{v^{\prime 2} m_{\phi}^3}{2 \pi v^4},  \  \  
  \Gamma (\phi^{++}\rightarrow W^+_T W^+_T) \approx  
  \frac{g^4v^{\prime 2}}{4\pi m_\phi },
  \nonumber
\end{eqnarray} 
where $W_L$ ($W_T$) stands for the longitudinal (transverse) component 
of the $W$ boson. We first point out that the $\Delta L=2$ processes,
$\phi^{++} \to \ell^+_i \ell^+_j$, are all driven by the
lepton number violating Yukawa coupling $Y_{ij}$. These decays to the lepton 
states will constitute the smoking gun signatures of the scenario
proposed by us. The decays into two gauge bosons, on the other 
hand, depend directly on $v^{\prime}$, the triplet vev.  The $m_{\phi}$
factors in the numerator in the 
decay to the longitudinally-polarized gauge bosons come from the typical
enhancement $(m_\phi^2/M_W^2)^2$ over the decay to the transversely-polarized
gauge bosons, governed by the Goldstone-boson equivalence theorem.
The $W_T^{\pm}W_T^{\pm}$ mode with a genuine gauge coupling 
thus becomes vanishingly small at higher $m_\phi$.

\begin{figure}[tb]
\includegraphics[height=3in,width=3in]{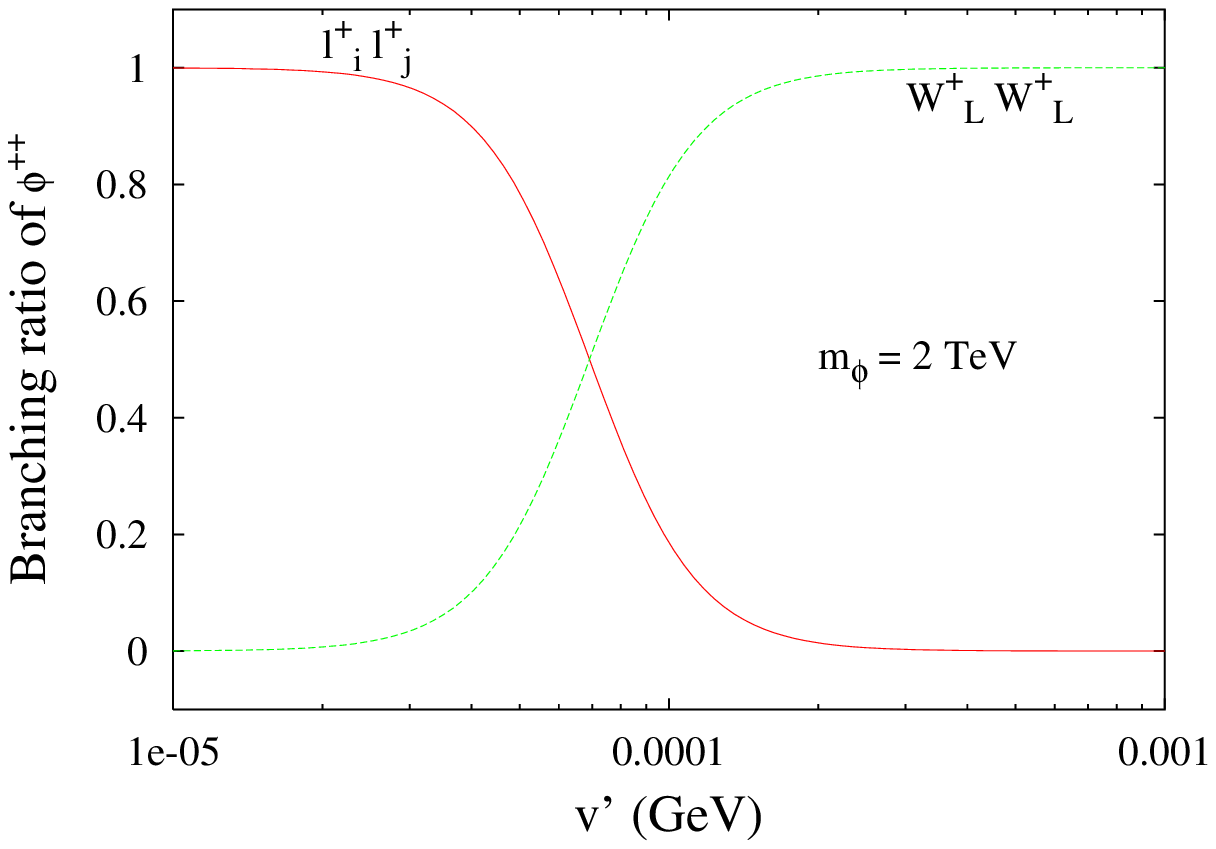}
\includegraphics[height=3in,width=3in]{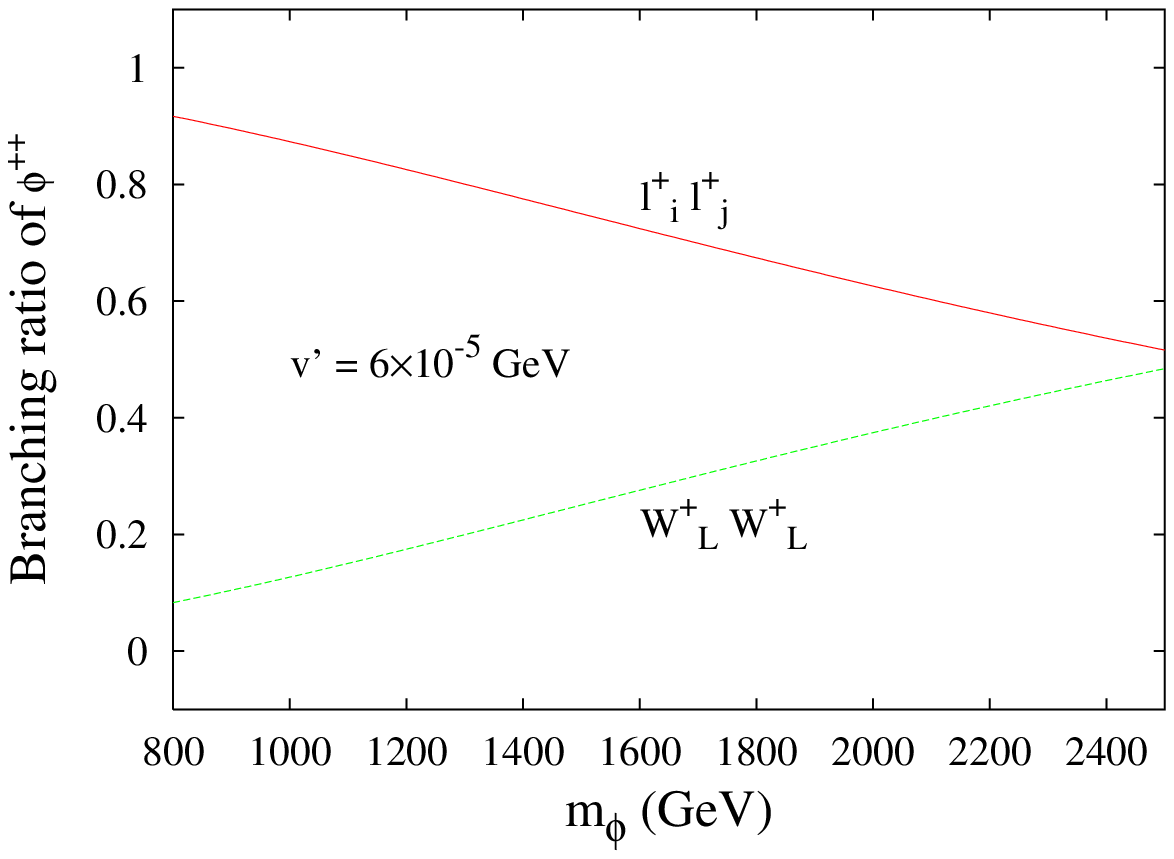}
\caption{Branching ratios of $\phi^{++}$ 
(a) versus the triplet vev for $Yv^{\prime}=10^{-10}$ GeV and 
$m_\phi =2$ TeV and 
(b) versus  $m_\phi$ for $v^{\prime}=6\times 10^{-5}$ GeV.
}
\label{fig:two}
\end{figure}

The complementarity between the $\ell^\pm \ell^\pm$ and $W^\pm W^\pm$
channels for small and large values of $v^{\prime}$ is clearly seen 
in Fig.~\ref{fig:two}:  for $m_{\phi} = 2$ TeV, 
the two channels are comparable 
when $v^{\prime} \approx 6\times 10^{-5}$.
In the calculation of the branching ratios of $\phi$ decays,
we sum over all six 
lepton flavor combinations in a flavor-democratic way
and we assume 
\begin{equation}
  Yv^{\prime} \approx 10^{-10}\ {\rm GeV} = 0.1\ {\rm eV},
  \label{eq:yv}
\end{equation}
so that neutrino masses lie in the expected range.  Note that for
$v^{\prime} \approx 6\times 10^{-5}$ GeV, this implies that 
$Y \approx 1.6\times 10^{-6}$.
While these couplings are still very small, we consider this parameter
freedom to be a strength of our analysis: our approach allows 
$Y \sim \mathcal{O}(1)$ with a very small $v^{\prime}$ but at the 
same time includes the possibility of small $Y$ as well, allowing a
large region of parameter space with interesting phenomenology.
We also present the branching ratio as a function of the $\phi^{++}$ mass
in Fig.~\ref{fig:two}(b) for $v^{\prime}=6\times 10^{-5}$ GeV.
Here one can  see the effect of the different $m_{\phi}$ dependence of 
the $\ell^+ \ell^+$ and $W^+ W^+$ final states.

It is interesting to note that the experimental data on neutrino mixing
require that at least some of the off-diagonal terms 
in $Y$ must be of the same order as the diagonal terms when written
in the charged lepton mass basis.  Although the
details of the structure depend on the particular neutrino
mass matrix, one can, assuming something like a flavor-democratic scenario,
immediately envision flavor violating decays such as 
$\phi^{\pm\pm} \to e^{\pm}\mu^{\pm}, \mu^{\pm}\tau^{\pm}$ of
sizable strength.  Such lepton flavor violating decays are a striking 
signal of this scanerio, where events with two like-sign different-flavor 
leptons can be observed in a decay final state which reconstructs
to an invariant mass peak at $m_\phi$.

\begin{figure}
\includegraphics[height=3in,width=3in]{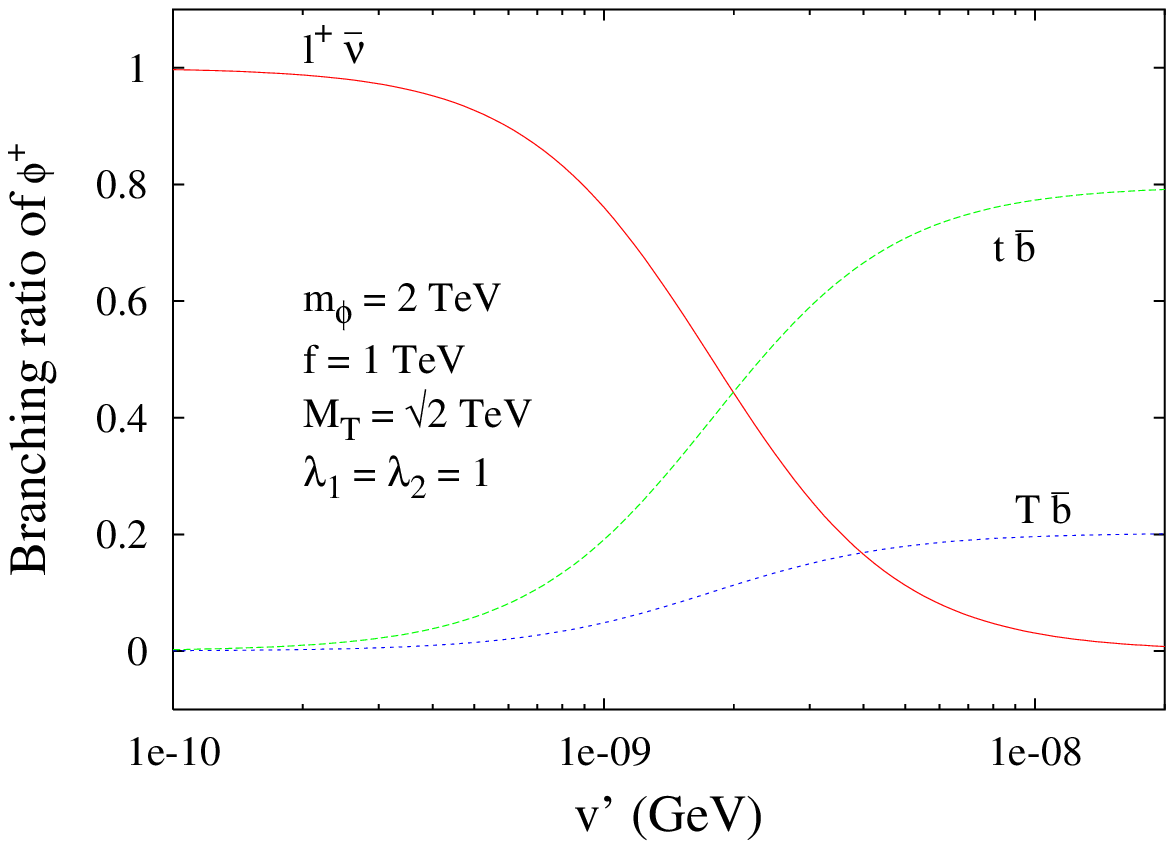}
\includegraphics[height=3in,width=3in]{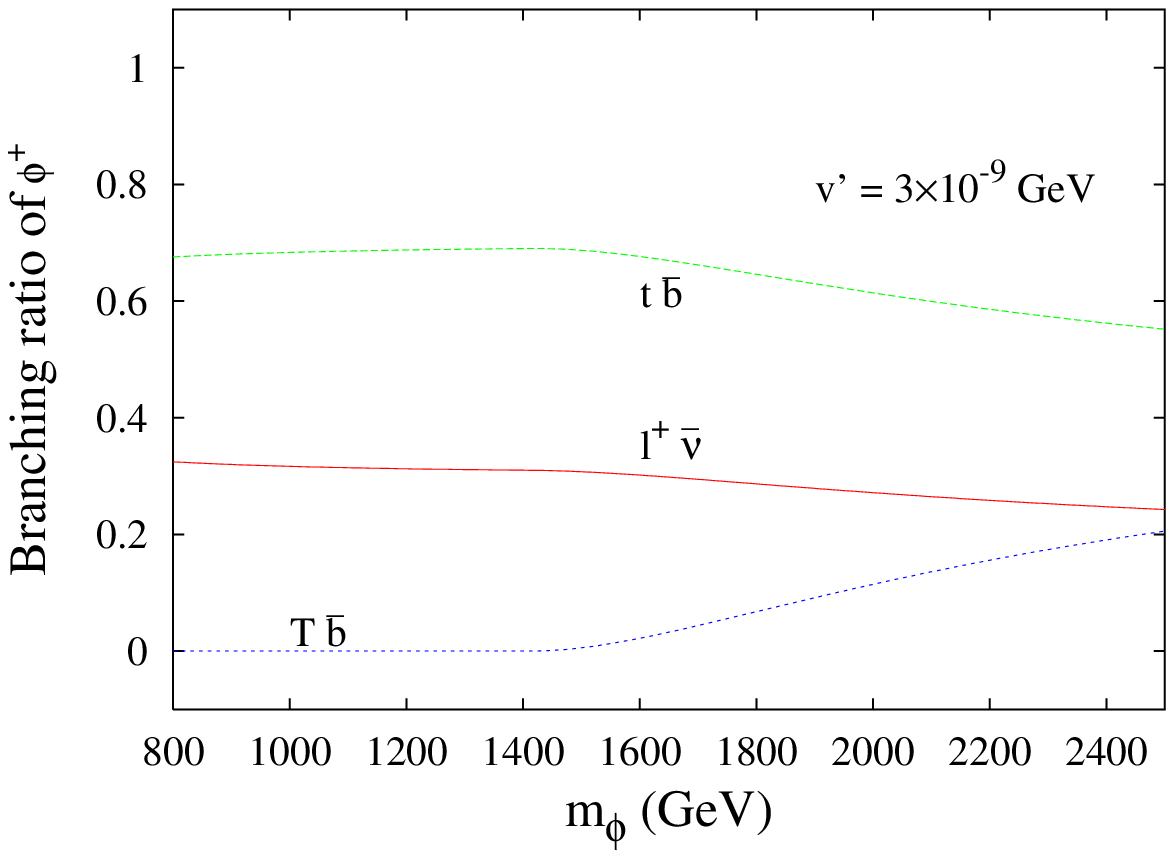}
\caption{Branching ratio of $\phi^{+}$ (a) versus the triplet vev
for $Yv^{\prime}=10^{-10}$ GeV and $m_\phi=2$ TeV and
(b) versus $m_{\phi}$ for $v^{\prime} = 3\times 10^{-9}$ GeV.
}
\label{fig:three}
\end{figure}

The branching ratios of $\phi^+$ and $\phi^0$ receive additional 
contributions from decays to heavy quarks.  Of course, an SU(2) triplet has no
dimension-four couplings to quarks.  However, in the LtH model
such couplings arise from ({\it i}) mixing between the triplet and the SU(2)
doublet Higgs at order $v^{\prime}/v$, and ({\it ii}) a dimension-five operator
involving both $H$ and $\phi$ that arises
from the expansion of the nonlinear sigma field in the top quark Yukawa
Lagrangian, Eq.~(\ref{eq:Lt}); inserting
the $H$ vev, this yields couplings of $\phi$ to heavy quark pairs 
suppressed by $v/f$.  Both of these contributions to the $\phi$ couplings
to heavy quarks are controlled by the relevant Yukawa couplings, $m_q/v$.
The two contributions, proportional to $v^{\prime}/v$ and $v/f$ respectively,
can be seen in the couplings given in Table~\ref{tab:feynman}.

We are interested in the parameter region $v/f \gg v^{\prime}/v$, in which
case the couplings of $\phi^+$ and $\phi^0$ to heavy quarks are
dominated by the dimension-five nonlinear sigma model operators, yielding
an interesting signal of the little Higgs structure in the top sector 
of the model.
Neglecting final-state masses, the partial decay widths are 
\begin{equation}
  \Gamma (\phi^{+}\rightarrow \ell^+_i \bar\nu_j) 
  = \frac{|Y_{ij}|^2 m_\phi}{8\pi }  ,\  \
  \Gamma (\phi^\pm \rightarrow t \bar{b},\bar{t} b) \approx
  \Gamma (\phi^s\rightarrow t \bar{t}) \approx 
  \Gamma (\phi^p\rightarrow t \bar{t}) \approx	
  \frac{N_c m^2_t}{16\pi f^2} m_\phi,
\end{equation}
where $N_c = 3$ is the number of colors.
The triplet couplings to $T\bar b$ and $T\bar t$ 
also involve the top sector parameters $\lambda_1$ and
$\lambda_2$ (see Appendix~\ref{app:A} for details) and the 
decay widths are proportional to $(\lambda_1/\lambda_2)^2$.  
We illustrate our results for $\lambda_1=\lambda_2$.  Exact formulae for the
partial widths are given in Appendix~\ref{app:B}.
Figures~\ref{fig:three} and \ref{fig:four} show that
the decays of $\phi^+$ and $\phi^0$ are dominated, 
approximately from $v^{\prime}=2\times 10^{-9}$ GeV upwards, 
by the heavy quark final states. 

Note that we have treated the triplet mass as a free parameter because of the 
arbitrary constants $a$ and $a^{\prime }$ in the coefficient of the 
triplet mass-squared, as explained 
in Appendix~\ref{app:A}. 
On the other hand, $M_T$ is proportional to $f$ for fixed 
$\lambda_1,\lambda_2$. Therefore a large value of $f$ in our approach,
while the free parameter $m_{\phi}$  is held fixed, will
suppress the decays into the $T$-quark. Our results are presented 
for $M_T = \sqrt{2}$ TeV. 

For $\phi^+$, the most interesting parameter range is where 
the elements of $Y$ range between $0.1$ and $1$, or equivalently
$v^{\prime}$ lies between $10^{-9}$ and $10^{-10}$ GeV.  In this case $\phi^+$ 
decays mostly into SM leptons, with branching fractions controlled
by the structure of the $Y_{ij}$ matrix, which of course directly
controls the neutrino masses and mixings.
The signatures of $\phi^+$ would then be quite distinct from
those of a charged scalar coming from a two-Higgs-doublet model, such
as in supersymmetric theories, in which the charged Higgs couplings to leptons
are directly proportional to the charged lepton masses.
It should also be remembered that this region, with $Y \sim \mathcal{O}(1)$, 
corresponds to the least number of fine-tuned parameters in the theory.
For larger values of $v^{\prime}$, however, the decays of
$\phi^+$ will be dominated by the heavy quark final states
$t\bar{b}$ (and $T\bar{b}$, if kinematically allowed) 
which are difficult to distinguish from
the decays of the charged Higgs of a two-Higgs-doublet model.
For $v^{\prime}$ below $10^{-4}$ GeV, the most distinct signals of the
triplet will be the $\phi^{\pm\pm}$ decays directly into like-sign dileptons.
It should be noted that the $\phi^{\pm\pm}$ does not have any hadronic
decay modes to compete with the $\Delta L = 2$ decays in this range of 
parameters.
For larger values of $v^{\prime}$, the
most distinct signals of the triplet will come from
$\phi^{\pm\pm}\rightarrow W^{\pm} W^{\pm}$, giving rise to
like-sign dileptons from the $W$ decays which can be identified with 
suitable event selection criteria.

\begin{figure}
\includegraphics[height=3in,width=3in]{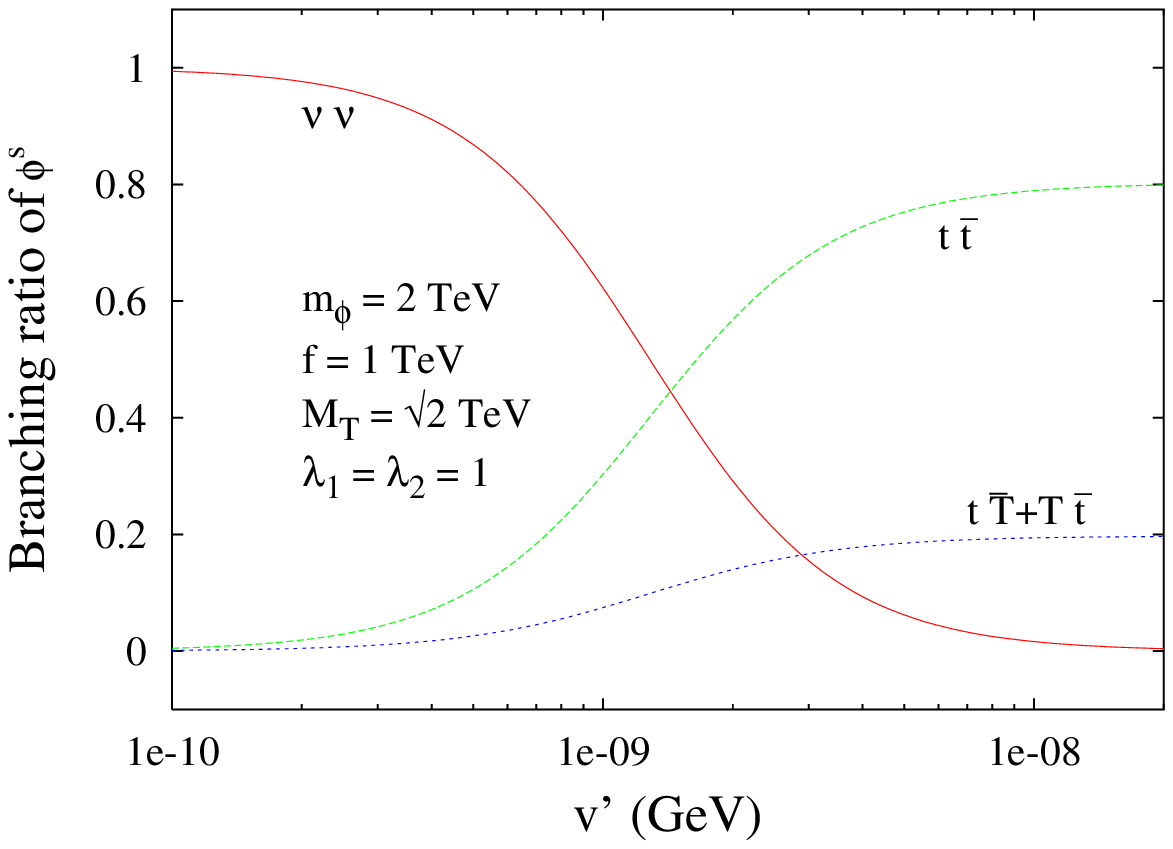}
\includegraphics[height=3in,width=3in]{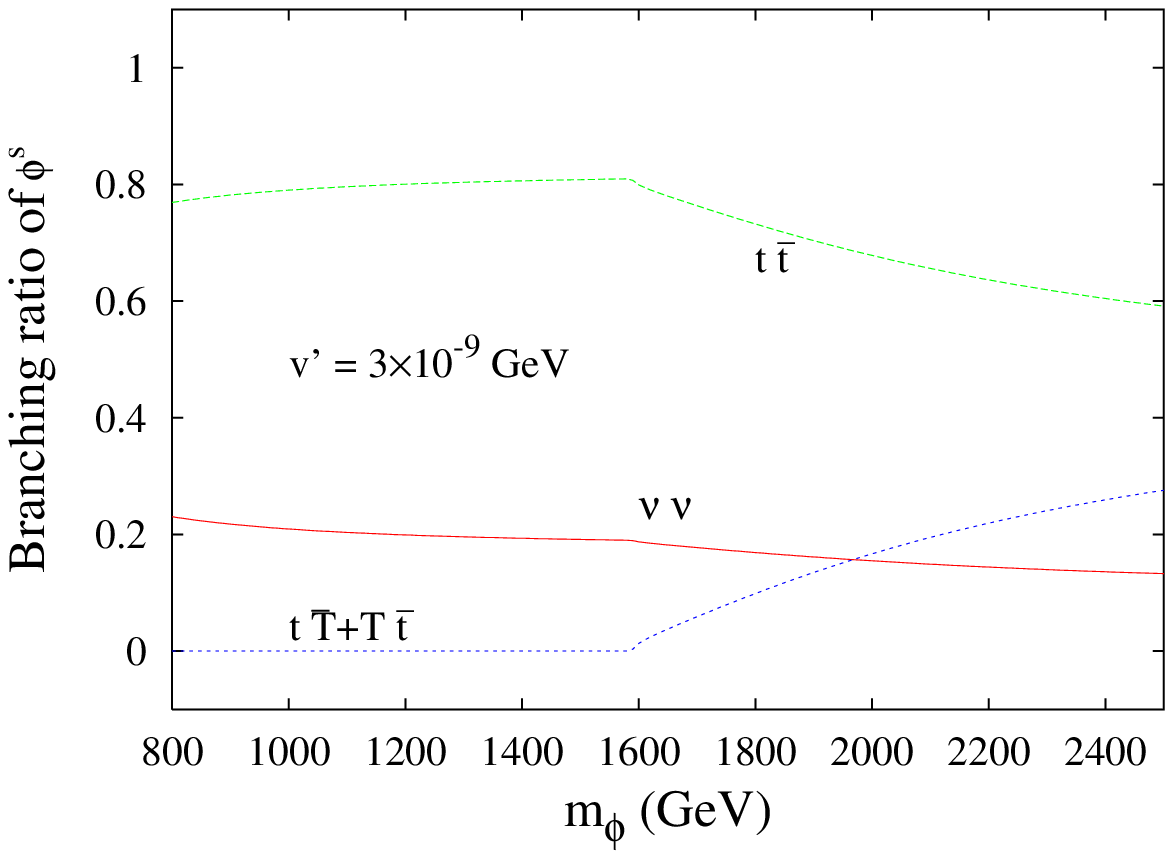}
\caption{Branching ratios of $\phi^{s}$ (a) versus the triplet vev
for $Yv^{\prime}=10^{-10}$ GeV and $m_\phi=2$ TeV,
and (b) versus $m_{\phi}$ for $v^{\prime} = 3\times 10^{-9}$ GeV.
The branching ratios of $\phi^p$ are virtually identical for the
parameter ranges shown.
}
\label{fig:four}
\end{figure}

In the same spirit, the neutral triplet states $\phi^s, \phi^p$ are
characterized by their invisble decays into two neutrinos for
$Y\gtrsim 0.1$, or equivalently $v^{\prime} \lesssim 10^{-9}$, as 
shown in Fig.~\ref{fig:four} for $\phi^s$.  The branching ratios of 
$\phi^p$ are virtually identical in this parameter range.
This makes the neutral scalar $\phi^s$ and the pseudoscalar $\phi^p$ 
quite different in appearance
from their counterparts in either the SM or a two-Higgs-doublet model.
Such invisible decays can lead to a detection of the neutral triplet 
through missing energy signatures 
or the identification of an invisible state recoiling against a $Z$ boson 
at a high-energy linear $e^{+}e^{-}$ collider. 

For the $\phi^{\pm}$, $\phi^s$ and $\phi^p$, the additional decay modes 
$\phi^{\pm}\rightarrow W^{\pm}h$, $\phi^s\rightarrow hh$, 
$\phi^p\rightarrow Zh$ are available with the same strength as the 
$W^{\pm}Z$ and $ZZ$ modes. However, all these channels are suppressed by 
$v^{\prime }/v$, and they do not stand a chance against either the heavy quark 
final states or the $\Delta L$ = 2 modes. Therefore, the production 
of the SM Higgs from triplet decays will be unobservable 
in this scenario. 

It should be noted that the region of the 
parameter space that gives
rise to these interesting signals involving leptons will not be
accessible in the scenario described in Sec.~\ref{sec:largersym} 
and Refs.~\cite{kilian,nuh1},
in which the $LL\phi$ operator is related to the dimension-five
$(LH)^2$ operator through the non-linear sigma model field.
Thus the decays of the triplet states can serve to distinguish 
between alternative scenarios for neutrino mass generation in the LtH model.

A final comment about the decay length of the triplets is in order here.
In the region where the $\ell^{\pm}\ell^{\pm}$ channel dominates, 
the lifetime $\tau$ of $\phi^{++}$ (with all flavours summed over) is given by
\begin{equation}
  \tau = \frac{8\pi }{9}\frac{v^{\prime 2}}{(Y_{ij}v^{\prime})^2} 
  \left(\frac{1\ {\rm TeV}}{m_\phi }\right) \times 6.6\times 10^{-28}\  {\rm sec}.
\end{equation}
For $Y_{ij}\approx 1.6\times 10^{-6}$
(or  $v^{\prime} \approx 6 \times 10^{-5}$ GeV), 
one finds $\tau \simeq 2.2 \times 10^{-16}$ sec for 
$m_{\phi}~=~2$ TeV. This gives 
a decay length $\ell_d \lesssim 0.1\ \mu$m, which is too short to show up as
a displaced vertex in the decay.  Taking a larger value for $v^{\prime}$
suppresses the partial width into like-sign lepton pairs, but the 
$WW$ mode then grows quickly and the decay length remains small.

\section{Summary and conclusions}
\label{four}

We have considered the simplest possible scenario for generating
the neutrino masses within the context of the Littlest Higgs model
by coupling the scalar triplet present
in the model to the leptons in a $\Delta L~=~2$ interaction.  This term
then generates neutrino masses through the triplet vev.
Although this term does not obey the overseeing [SU(2)$\times$U(1)]$^2$ 
gauge invariance, 
it does not affect the cancellation of quadratic divergences in the 
Higgs mass.
We also showed that all contributions coming from
dimension-five operators remain subdominant so long as one assumes
that there is no
lepton-number violating new physics at the scale $\Lambda$. Following
the phenomenological requirement of 
keeping the neutrino masses in the required range, we are led to a situation
where either the lepton number violating Yukawa coupling or the triplet 
vev has to be very small.  The second possibility, presumbaly triggered 
by some yet-unknown 
feature of the Coleman-Weinberg effective potential, allows one to retain the
lepton number violating couplings to be $\mathcal{O}(1)$, 
a situation that seems less fine-tuned from the viewpoint of allowing 
bi-large mixing in the neutrino sector. 

We have also investigated the decays of the triplet scalar states in this
scenario and identified their characteristic features associated
with lepton number violation. 
The most striking signature comes from 
the doubly-charged scalar decays. The crucial test is 
the complementarity between the final states of
$W^\pm W^\pm$ and $\ell^\pm \ell^\pm$:
While the triplet vev controls the $W^\pm W^\pm$ mode and 
thus the final state branching ratios 
over a large range, the region corresponding
to $Y\approx 1$ leads to significant $\Delta L=2$ modes,
with possibly large lepton-flavor violation.
Different complementarity exists for the other triplet scalar decays:
between SM heavy quarks (independent of $v^{\prime}$) and the 
$\Delta L=2$ modes. 
Moreover, the singly-charged scalar may decay to charged leptons
with nearly universal couplings, unlike the charged Higgs in typical 
two-Higgs-doublet models. Another interesting consequence is the 
``invisible'' decay of the neutral triplet state into two neutrinos.
These decays would allow one to distinguish models of lepton flavor 
violation within the Littlest Higgs scenario and directly constrain
the elements of the $\Delta L=2$ coupling matrix which controls the 
neutrino masses and mixings.

\begin{acknowledgments}
We thank Bob McElrath and Liantao Wang for
numerous discussions about the neutrino mass issues in the LH scenarios.
BM thanks the hospitality of  the Phenomenology
Institute at the University of Wisconsin--Madison, where this work was 
initiated.  TH would like to thank the CERN Theory Division for the hospitality
during the final stage of this work.
TH and HEL were supported in part by the U.S.~Department of 
Energy under grant DE-FG02-95ER40896 and in part by the Wisconsin 
Alumni Research Foundation.
\end{acknowledgments}

\appendix
\section{The Littlest Higgs model}
\label{app:A}

\subsection{Brief summary of the LtH model}
\label{app:A1}

The little Higgs approach conceives the Higgs boson as member of a set 
of pseudo-Goldstone bosons.
In the original version of the 
Littlest Higgs (LtH) scenario \cite{Nima} to be discussed here, the 
pseudo-Goldstone bosons arise when a global SU(5) symmetry is broken 
down to SO(5) at a scale $\Lambda \sim 4\pi f$. 
These pseudo-Goldstone bosons are described by a nonlinear sigma model below 
the scale $\Lambda$. 

The breakdown of the global symmetry is triggered by a vacuum expectation 
value (vev) $\Sigma_0$ of the sigma-model field,
\begin{equation}
	\Sigma = e^{i\Pi /f}\Sigma _0e^{i\Pi ^T/f},
\end{equation}
where $\Pi = \sum_a\Pi^aX^a$ and $X^a$ correspond to the 14
broken SU(5) generators.  Explicitly, we have
\begin{equation}
\Sigma_0 = \left( \begin{array}{ccc}
           & & {\bf 1}_{2\times 2} \\
	   & 1 & \\
	   {\bf 1}_{2\times 2} & & 
	   \end{array} \right), 
	\qquad \qquad
\Pi = \left( \begin{array}{ccc}
   {\bf 0}_{2\times 2} & \frac {H^\dag }{\sqrt2} & \phi ^\dag \\
   \frac {H^* }{\sqrt2} & 0 & \frac {H}{\sqrt2}\\
   \phi  & \frac {H^T }{\sqrt2} & {\bf 0}_{2\times 2}
   \end{array} \right),
\end{equation}
where we have suppressed the Goldstone modes that will later be eaten
by broken gauge generators, and we define
\begin{equation}
H = (h^+,h^0),
	\qquad
\phi = {-i} \left( \begin{array}{cc}
	\phi^{++} & \frac{\phi^+}{\sqrt2} \\
       \frac{\phi^+}{\sqrt{2}} & \phi^0 
	\end{array} \right).
	\label{eq:Hphixi}
\end{equation}
An [SU(2)$\times$U(1)]$^2$ subgroup of the global SU(5) is gauged.
The $\Sigma_0$ vev that is responsible for the breakdown of the
global symmetry also breaks the gauged [SU(2)$\times$U(1)]$^2$ 
down to the SM electroweak gauge symmetry SU(2)$_L\times$U(1)$_Y$.
Under the electroweak gauge group,
$H$ and $\phi$ transform as a complex doublet and 
a complex triplet, respectively. 

The gauge interaction of the sigma field is encoded in its covariant
derivative:
\begin{equation}
	{\cal L}_\Sigma = \frac{f^2}{8}Tr|D_{\mu}\Sigma|^2,
\end{equation}
where
\begin{equation} 
	D_{\mu}\Sigma = \partial_{\mu}\Sigma -i\sum_{j=1,2}
	[g_jW_{j\mu }^a(Q_j^a\Sigma + \Sigma Q_j^{aT}) 
	+ g^{\prime}_jB_{j\mu }(Y_j\Sigma + \Sigma Y_j^T)].
\end{equation}
Here $Q_j^a$ are the SU(2) generators and $Y_j$ are the U(1) generators,
which explicitly break the global SU(5) symmetry:
\begin{equation}
Q_1^a = \left( \begin{array}{cc}
	\frac{\sigma^a}{2} & \\
	    & {\bf 0}_{3\times 3}
	\end{array} \right),
\qquad \qquad
Q_2^a = \left( \begin{array}{cc}
	{\bf 0}_{3\times 3} & \\
	    & \frac{\sigma^{a*}}{2}
	\end{array} \right),
\end{equation}
\begin{equation}
	Y_1 = \frac{1}{10} {\rm diag}(-3,-3,2,2,2),
	\qquad \qquad
	Y_2 = \frac{1}{10} {\rm diag}(-2,-2,-2,3,3).
\end{equation}
Notice that setting $g_1 = g_1^{\prime} = 0$ leaves unbroken an SU(3)
subgroup of the global SU(5) symmetry; we call this remaining global symmetry
SU(3)$_1$.  Similarly, setting
$g_2 = g_2^{\prime} = 0$ leaves unbroken a second SU(3)
subgroup of the global SU(5) symmetry, which we call SU(3)$_2$.
The Higgs doublet $H$ transforms nonlinearly under both of these global
SU(3) symmetries, and thus remains an exact Goldstone boson so long as
these global symmetries are not explicitly broken.  A Higgs mass term can
thus be generated only by interactions involving both $g_1$ and $g_2$ 
(or both $g_1^{\prime}$ and $g_2^{\prime}$); this serves to forbid the
diagrams that generate the quadratic divergence in the Higgs mass at 
one loop.
However, logarithmically divergent 
diagrams contributing to the Higgs mass at one loop involve both gauge 
couplings $g_1$ and $g_2$ (or both $g_1^{\prime}$ and $g_2^{\prime}$)
and thus break the global SU(3), thereby leading to contributions to 
the Higgs mass.

In order to cancel the quadratic divergence arising through the top quark 
Yukawa coupling, we have to 
introduce a heavy vector-like quark pair 
($T,T^c$), where $T$ is left-handed and has charge +2/3. 
Including this vectorlike pair, the top Yukawa Lagrangian is
\begin{equation}
	{\cal L}_t = \frac{\lambda_1}{2}f\epsilon_{ijk}\epsilon_{xy}\chi_i
	\Sigma_{jx}\Sigma_{ky}t^c + \lambda_2fT T^c
	+ {\rm h.c.},
	\label{eq:Lt}
\end{equation}
where $\chi^T = (b_L,t_L,T)$ and $t^c$ is an SU(2) singlet. The 
indices $i,j,k$ take the values 1,2,3, whereas $x,y$ take the values 4,5. 
It should be noted here that the coupling $\lambda_1$ preserves the global
SU(3)$_1$ and breaks SU(3)$_2$, while $\lambda_2$ preserves SU(3)$_2$
and breaks SU(3)$_1$.
This ensures that the Higgs mass-squared is protected from quadratic 
divergences involving the top quark sector at one loop.
Diagonalizing the mass matrix arising from Eq.~(\ref{eq:Lt}), we find the 
physical top quark $t$ and a heavy isospin-singlet ``top-partner'' $T$:
\begin{equation}
  m_t \simeq \frac{\lambda_1 \lambda_2 }{\sqrt{\lambda^2_1+\lambda^2_2}}v,
	\qquad \qquad
  M_T \simeq f\sqrt{\lambda^2_1+\lambda^2_2}.
\end{equation}

The gauge and top quark interactions generate a Higgs potential at one loop
via the Coleman-Weinberg mechanism \cite{cw}, which is given by 
\begin{eqnarray}
	V_{CW} = \lambda_{\phi^2}f^2{\rm Tr}(\phi^\dag \phi) 
	+ i\lambda_{h\phi h}f(H\phi^\dag H^T-H^*\phi H^\dag) - \mu^2HH^\dag 
	+ \lambda_{h^4}(HH^\dag)^2
	\nonumber \\
	+ \lambda_{h\phi \phi h}H\phi^\dag\phi H^\dag 
	+ \lambda_{h^2\phi^2}HH^\dag {\rm Tr}(\phi^\dag \phi ) 
	+ \lambda_{\phi^2\phi^2}[{\rm Tr}(\phi^\dag \phi )]^2 
	+ \lambda_{\phi^4}{\rm Tr}(\phi^\dag \phi \phi^\dag \phi ),
\end{eqnarray}
with coefficients
\begin{eqnarray}
	\lambda_{\phi^2}& =& 
	\frac{a}{2}\left[\frac{g^2}{s^2c^2}
	+\frac{g^{\prime 2}}{s^{\prime 2}c^{\prime 2}}\right] 
	+ 8a^{\prime}\lambda_1^2
	\label{lambda2} \\
	\lambda_{h\phi h} &=& -\frac{a}{4}\left[g^2\frac{(c^2-s^2)}{s^2c^2}
	+g^{\prime 2}\frac{(c^{\prime 2}-s^{\prime 2})}
		{s^{\prime 2}c^{\prime 2}}\right] 
	+ 4a^{\prime}\lambda_1^2\\
	\lambda_{h^4} &=& \frac{1}{4}\lambda_{\phi^2},
	\qquad
	\lambda_{h\phi \phi h} = -\frac{4}{3}\lambda_{\phi^2},
	\qquad
	\lambda_{\phi^2\phi^2} = -16a^{\prime}\lambda_1^2\\
	\lambda_{\phi^4} &=& -\frac{2a}{3}\left[\frac{g^2}{s^2c^2}
	+\frac{g^{\prime 2}}{s^{\prime 2}c^{\prime 2}}\right] 
	+ \frac{16a^{\prime}}{3}\lambda_1^2.
\end{eqnarray} 
where $c$ and $s$ ($c^\prime$ and $s^\prime$) are the gauge coupling
mixing parameters for the SU(2) (U(1)) gauge groups,
respectively \cite{han}.
Here $a$, $a^{\prime}$ are parameters of $\mathcal{O}(1)$ that encapsulate
the cutoff dependence of the gauge and top sectors, respectively, of the
UV-incomplete theory.
The parameters $\mu^2$ and $\lambda_{h^2\phi^2}$ are generated through 
logarithmic contributions. 
Electroweak symmetry breaking is triggered if $\mu^2>0$, whereby the scalar 
doublet acquires a vev. The triplet vev is kept small by keeping 
$\lambda_{\phi^2}$ positive;
it originates in mixing with the doublet via $\lambda_{h \phi h}$. 
The minimization conditions 
for $V_{CW}$, in terms of $\langle h^0 \rangle = v/\sqrt{2}$, 
$\langle \phi^0 \rangle = v^{\prime}$, are 
\begin{eqnarray}
	v^2 = \frac{\mu^2}{\lambda_{h^4}
	-\frac{\lambda_{h\phi h}^2}{\lambda_{\phi^2}}}, 
	\qquad \qquad
	v^{\prime} = \frac{\lambda_{h\phi h}v^2}{2\lambda_{\phi^2}f}.
	\label{eq:vvprime}
\end{eqnarray}
Note that terms of the form $H^2\phi^2,\phi^4$ give
a subleading contribution to Eq.~(\ref{eq:vvprime}) and 
have been neglected. In order to ensure electroweak symmetry
breaking, we should have 
$\lambda_{h^4}-\frac{\lambda_{h\phi h}^2}{\lambda_{\phi^2}}>0$.
The resulting masses for the triplet states $\phi$ and the physical Higgs
boson $h$ after electroweak symmetry breaking are
\begin{eqnarray}
  m^2_\phi \simeq \lambda_{\phi^2}f^2,
  \qquad \qquad
  m^2_h \simeq 2 \left( \lambda_{h^4} 
	 -\frac{\lambda_{h\phi h}^2}{\lambda_{\phi^2}} \right)
               v^2 \simeq 2\mu^2.
\end{eqnarray}

It should also be noted that $\lambda_{\phi^2}$, as expressed above, gets 
modified by an additional term once $\Delta L = 2$ interactions are 
switched on, as has been shown in Sec.~\ref{sec:lphil}.

\subsection{Lepton number violation}

\begin{table}
\begin{center}
\begin{tabular}{|c|c|}
\hline \hline
$\phi^{--} \ell^+_i \ell^+_j \ \ (i \leq j)$ &
        $2 i Y_{ij}^* P_R C$ \\

$\phi^- \ell_i^+ \bar \nu_j$ &
        $i \sqrt{2} Y_{ij}^* P_R C$ \\

$\phi^s \nu_i \nu_j \ \ (i \leq j)$ &
        $i \sqrt{2} Y_{ij} C^{-1} P_L$ \\
$\phi^s \bar \nu_i \bar \nu_j \ \ (i \leq j)$ &
        $i \sqrt{2} Y_{ij}^* P_R C$ \\

$\phi^p \nu_i \nu_j \ \ (i \leq j)$ &
        $- \sqrt{2} Y_{ij} C^{-1} P_L$ \\
$\phi^p \bar \nu_i \bar \nu_j \ \ (i \leq j)$ &
        $\sqrt{2} Y_{ij}^* P_R C$ \\
\hline \hline
\end{tabular}
\end{center}
\caption{ Feynman rules for $\Delta L$ = 2 couplings. All particles and 
momenta are outgoing. $C$ is the charge-conjugation operator. 
Since $Y_{ij}$ is symmetric under $(i,j)$ we have combined the symmetric
vertices involving $\phi^{--}$, $\phi^s$ and $\phi^p$ 
and written them only for $i\le j$.}
\label{tab:feynmanDL2}
\end{table}

When we introduce the $\Delta L=2$ interaction  of 
Eq.~(\ref{lphil}) in order to give rise to neutrino masses, 
one of its effects  is
to add an extra term to the expression of Eq.~(\ref{lambda2}) 
for $\lambda_{\phi^2}$, as shown in Eq.~(\ref{eq:lambdaphi2}).
This contribution is typically small in the parameter ranges 
that we consider.

As for the $\Delta L=2$  interactions of the triplet $\phi$, 
expanding Eq.~(\ref{lphil}) explicitly one can obtain the full lepton number
violating interaction vertices. The dimension-four couplings are 
given in Eq.~(\ref{app:lphil}).
The Feynman rules for the $\Delta L = 2$ interactions are given
in Table~\ref{tab:feynmanDL2}.
The relevant lepton number conserving interactions between 
the triplet state and SM particles \cite{han} are given as Feynman
rules in Table~\ref{tab:feynman}.  For the $\phi^s hh$ coupling, we
have included the symmetry factor, Feynman rule $= i \mathcal{L} \times 2$,
and used the relation in Eq.~(\ref{eq:vvprime}) to write $\lambda_{h\phi h}$
in terms of $v^{\prime}$.

\begin{table}
\begin{center}
\begin{tabular}{|c|c|}
\hline \hline
$\phi^{--} W^+_{\mu} W^+_{\nu}$ &
	$2 i g^2 v^{\prime} g_{\mu \nu}$ \\
\hline
$\phi^- W^+_{\mu} Z_{\nu}$ &
	$-i \frac{g^2}{c_W} v^{\prime} g_{\mu \nu}$ \\
$\phi^- W^+_{\mu} h$ &
	$-i g \frac{v^{\prime}}{v} (p_h - p_{\phi})_{\mu}$ \\
$\phi^- \bar b t$ &
	$-\frac{i}{\sqrt{2}v} (m_t P_R + m_b P_L) 
	(\frac{v}{f} - 4 \frac{v^{\prime}}{v})$  \\
$\phi^- \bar b T$ &
	$-\frac{im_t}{\sqrt{2}v} (\frac{v}{f} - 4 \frac{v^{\prime}}{v})
	\frac{\lambda_1}{\lambda_2} P_R$ \\
\hline
$\phi^s Z_{\mu} Z_{\nu}$ &
	$i \sqrt{2} \frac{g^2}{c_W^2} v^{\prime} g_{\mu \nu}$ \\
$\phi^s h h$ &
	$i 2 \sqrt{2} m_{\phi}^2 \frac{v^{\prime}}{v^2}$ \\
$\phi^s W^+_{\mu} W^-_{\nu}$ &
	$0$ \\
$\phi^s \bar t t$ &
	$-\frac{im_t}{\sqrt{2}v} (\frac{v}{f} - 4 \frac{v^{\prime}}{v})$ \\
$\phi^s \bar b b$ & 
	$-\frac{im_b}{\sqrt{2}v} (\frac{v}{f} - 4 \frac{v^{\prime}}{v})$ \\
$\phi^s \bar t T$ &
	$-\frac{im_t}{\sqrt{2}v} (\frac{v}{f} - 4 \frac{v^{\prime}}{v})
	\frac{\lambda_1}{\lambda_2} P_R$ \\
$\phi^s \bar T t$ &
	$-\frac{im_t}{\sqrt{2}v} (\frac{v}{f} - 4 \frac{v^{\prime}}{v})
	\frac{\lambda_1}{\lambda_2} P_L$ \\
\hline
$\phi^p Z_{\mu} h$ &
	$-\sqrt{2} \frac{g}{c_W} \frac{v^{\prime}}{v} 
	(p_h - p_{\phi})_{\mu}$ \\
$\phi^p \bar t t$ &
	$-\frac{m_t}{\sqrt{2}v} (\frac{v}{f} - 4 \frac{v^{\prime}}{v}) 
	\gamma^5$ \\
$\phi^p \bar b b$ &
	$\frac{m_b}{\sqrt{2}v} (\frac{v}{f} - 4 \frac{v^{\prime}}{v}) 
	\gamma^5$ \\
$\phi^p \bar t T$ &
	$\frac{m_t}{\sqrt{2}v} (\frac{v}{f} - 4 \frac{v^{\prime}}{v})
	\frac{\lambda_1}{\lambda_2} P_R$ \\
$\phi^p \bar T t$ &
	$\frac{m_t}{\sqrt{2}v} (\frac{v}{f} - 4 \frac{v^{\prime}}{v})
	\frac{\lambda_1}{\lambda_2} P_L$ \\
\hline \hline
\end{tabular}
\end{center}
\caption{Feynman rules for lepton number conserving $\phi$ couplings 
to SM particles, from Ref.~\cite{han}.  All particles and momenta are
outgoing.
}
\label{tab:feynman}
\end{table}

\section{Triplet decay partial widths}
\label{app:B}

In this Appendix we present the formulas for the triplet decay partial widths.
We define the standard kinematic function 
$\lambda (x,y,z) = x^2 + y^2 + z^2 - 2xy - 2xz - 2yz$ and use the scaled
mass variable $r_i = m_i/m_\phi$.  For the doubly-charged scalar $\phi^{++}$,
we have
\begin{eqnarray}
	\Gamma (\phi^{++}\rightarrow \ell^+_i \ell^+_j) &=& 
	\left\{ \begin{array}{lr}
	\frac{1}{8\pi } |Y_{ij}|^2 m_\phi, & \qquad (i=j) \\
	\frac{1}{4\pi }|Y_{ij}|^2 m_\phi, & \qquad (i<j) \end{array} \right.
		 \nonumber \\
	\Gamma (\phi^{++}\rightarrow W^+_T W^+_T) &=& \frac{1}{4\pi }
	\frac{g^4v^{\prime 2}}{m_\phi }
	\frac{\lambda^{\frac{1}{2}}(1 ,r^2_W ,r^2_W)}{\sqrt{4r_W^2 + 
	\lambda (1 ,r^2_W ,r^2_W)}} \approx  
	\frac{g^4v^{\prime 2}}{4\pi m_\phi }, 
		\nonumber \\
	\Gamma (\phi^{++}\rightarrow W^+_L W^+_L) &=& \frac{1}{4\pi }
	\frac{g^4v^{\prime 2}}{2m_\phi }
	\frac{\lambda^{\frac{1}{2}}(1 ,r^2_W ,r^2_W)}{\sqrt{4r_W^2 + 
	\lambda (1 ,r^2_W ,r^2_W)}} \frac{(1 - 4r_W^2)^2}{4r_W^4} 
	\approx  \frac{v^{\prime 2} m_{\phi}^3}{2 \pi v^4},
\end{eqnarray}
where in the last two expressions we have shown the approximate result
neglecting final-state masses compared to $m_{\phi}$.
We use the subscripts $T$ and $L$ to denote the transverse and longitudinal 
polarizations of the SM gauge bosons.

For the singly-charged scalar $\phi^+$, we have,
\begin{eqnarray}
	\Gamma (\phi^+\rightarrow \ell^+_i \bar\nu_j) &=& \frac{1}{8\pi }
	|Y_{ij}|^2 m_\phi,
		\nonumber \\
	\Gamma (\phi^+\rightarrow W_T^+ Z_T) &=&  \frac{1}{4\pi }
	\frac{g^4v'^2}{m_\phi c^2_W}
	\left[\frac{\lambda^{\frac{1}{2}}(1,r^2_W ,r^2_Z)}
		{\sqrt{4r_W^2 + \lambda (1,r^2_W ,r^2_Z)} 
		+ \sqrt{4r_Z^2 + \lambda (1,r^2_W ,r^2_Z)}}\right]
		\nonumber \\
	&\approx & \frac{g^4 v^{\prime 2} }{8 \pi m_\phi c^2_W} ,
		\nonumber \\
	\Gamma (\phi^+\rightarrow W_L^+ h) &=& \frac{1}{4\pi }
	\frac{g^2v'^2}{v^2} \frac{m_\phi}{2r_W^2}
	 \left[\frac{\lambda^{\frac{3}{2}}(1,r^2_h ,r^2_W)}
		{\sqrt{4r_h^2 + \lambda (1,r^2_h ,r^2_W)} 
		+ \sqrt{4r_W^2 + \lambda (1,r^2_h ,r^2_W)}}\right]
		\nonumber \\
	&\approx &\frac{v^{\prime 2}m^3_\phi }{4\pi v^4} 
		\nonumber \\
	\Gamma (\phi^+\rightarrow W_L^+ Z_L) &=&  \frac{1}{4\pi }
	\frac{g^4v'^2}{2m_\phi c^2_W}
	\left[\frac{\lambda^{\frac{1}{2}}(1,r^2_W ,r^2_Z)}
		{\sqrt{4r_W^2 + \lambda (1,r^2_W ,r^2_Z)} 
		+ \sqrt{4r_Z^2 + \lambda (1,r^2_W ,r^2_Z)}}\right]
		\nonumber \\
	&& \times \frac{(1 - r_W^2 - r_Z^2)^2}{4 r_W^2 r_Z^2}
	\approx  \frac{v^{\prime 2} m_{\phi}^3}{4 \pi v^4}
		\nonumber \\
	\Gamma (\phi^+\rightarrow t \bar{b}) &=& \frac{N_c}{4\pi }
	\frac{r^2_t m^3_\phi }{4f^2}
	\left[\frac{\lambda^{\frac{1}{2}}(1,r^2_t ,r^2_b) (1 - r_t^2 - r_b^2)}
	{\sqrt{4r_t^2 + \lambda (1,r^2_t ,r^2_b)} 
	+ \sqrt{4r_b^2 + \lambda (1,r^2_t ,r^2_b)}}\right]
	\approx  \frac{N_c m^2_t m_\phi }{32 \pi f^2},
		\nonumber \\
	\Gamma (\phi^+\rightarrow T \bar{b}) &=& \frac{N_c}{4\pi }
	\frac{r^2_t m^3_\phi }{4f^2} 
	\left[\frac{\lambda^{\frac{1}{2}}(1,r^2_T ,r^2_b) (1 - r_T^2 - r_b^2)}
	{\sqrt{4r_T^2 + \lambda (1,r^2_T ,r^2_b)} 
	+ \sqrt{4r_b^2 + \lambda (1,r^2_T ,r^2_b)}}\right]
	\left(\frac{\lambda_1}{\lambda_2}\right)^2
			\nonumber \\
	&\approx & \frac{N_c m^2_t m_\phi }{32 \pi f^2}
		\left(\frac{\lambda_1}{\lambda_2}\right)^2
		(1-r^2_T)^2.
\end{eqnarray}

For the neutral scalar $\phi^s$, we have
\begin{eqnarray}
	\Gamma (\phi^s\rightarrow \nu_i \nu_j + \bar \nu_i \bar \nu_j ) &=& 
	\left\{ \begin{array}{lr}
	\frac{1}{8\pi } |Y_{ij}|^2 m_\phi, & \qquad (i=j) \\
	\frac{1}{4\pi } |Y_{ij}|^2 m_\phi, & \qquad (i<j) \end{array} \right.
		\nonumber \\
	\Gamma (\phi^s\rightarrow Z_T Z_T) &=&  \frac{1}{4\pi }
	\frac{g^4v'^2}{2m_\phi c^4_W}
	\frac{\lambda^{\frac{1}{2}}(1,r^2_Z ,r^2_Z)}
	{\sqrt{4r_Z^2 + \lambda (1,r^2_Z ,r^2_Z)}}
	\approx  \frac{g^4v'^2}{8\pi m_\phi c^4_W}
		\nonumber \\
	\Gamma (\phi^s\rightarrow h h) &=&  \frac{1}{4\pi }
	\frac{v^{\prime 2}m^3_\phi }{v^4}
	\frac{\lambda^{\frac{1}{2}}(1,r^2_h ,r^2_h)}
	{\sqrt{4r_h^2 + \lambda (1,r^2_h ,r^2_h)}}
	\approx  \frac{v^{\prime 2}m^3_\phi }{4\pi v^4}
		\nonumber \\
	\Gamma (\phi^s\rightarrow Z_L Z_L) &=& \frac{1}{4\pi }
	\frac{g^4v'^2}{4m_\phi c^4_W}
	\frac{\lambda^{\frac{1}{2}}(1,r^2_Z ,r^2_Z)}
	{\sqrt{4r_Z^2 + \lambda (1,r^2_Z ,r^2_Z)}}
	\frac{(1 - 4r_Z^2)^2}{4r_Z^4} 
	\approx \frac{v^{\prime 2} m_{\phi}^3}{4 \pi v^4}
		\nonumber \\ 
	\Gamma (\phi^s\rightarrow t \bar{t})& =& \frac{N_c}{4\pi }
	\frac{r^2_t m^3_\phi }{4f^2}
	\frac{\lambda^{\frac{1}{2}}(1,r^2_t ,r^2_t)}
	{\sqrt{4r_t^2 + \lambda (1,r^2_t ,r^2_t)}}(1 - 4r_t^2)
	\approx  \frac{N_c m^2_t m_\phi }{16\pi f^2},
		\nonumber \\
	\Gamma (\phi^s\rightarrow b \bar{b}) &=& \frac{N_c}{4\pi }
	\frac{r^2_bm^3_\phi }{4f^2}
	\frac{\lambda^{\frac{1}{2}}(1,r^2_b ,r^2_b)}
	{\sqrt{4r_b^2 + \lambda (1,r^2_b ,r^2_b)}}(1 - 4r_b^2) 
	\approx \frac{N_c m^2_b m_\phi }{16\pi f^2},
		\nonumber \\
	\Gamma (\phi^s\rightarrow T \bar{t} + t \bar{T}) &=& \frac{N_c}{4\pi }
	\frac{r^2_tm^3_\phi }{2f^2}
	\left[\frac{\lambda^{\frac{1}{2}}(1,r^2_T ,r^2_t) (1 - r_T^2 - r_t^2)}
	{\sqrt{4r_T^2 + \lambda (1,r^2_T ,r^2_t)} 
	+ \sqrt{4r_t^2 + \lambda (1,r^2_T ,r^2_t)}}\right]
	\left(\frac{\lambda_1}{\lambda_2}\right)^2
				\nonumber \\
	&\approx & \frac{N_c m^2_t m_\phi }{16 \pi f^2}
		\left(\frac{\lambda_1}{\lambda_2}\right)^2
		(1-r^2_T)^2.
\end{eqnarray}

Finally, for the neutral pseudoscalar $\phi^p$, we have
\begin{eqnarray}
	\Gamma (\phi^p\rightarrow \nu_i \nu_j  + \bar \nu_i \bar \nu_j) &=& 
	\left\{ \begin{array}{lr}
	\frac{1}{8\pi } |Y_{ij}|^2 m_\phi, & \qquad (i=j) \\
	\frac{1}{4\pi } |Y_{ij}|^2 m_\phi, & \qquad (i<j) \end{array} \right.
		\nonumber \\
	\Gamma (\phi^p\rightarrow Z_L h) &=&  \frac{1}{4\pi }
	\frac{g^2 v^{\prime 2} m_\phi }{v^2 c^2_W r^2_Z}
	\left[\frac{\lambda^{\frac{3}{2}}(1,r^2_h ,r^2_Z)}
	{\sqrt{4r_h^2 + \lambda (1,r^2_h ,r^2_Z)} 
	+ \sqrt{4r_Z^2 + \lambda (1,r^2_h ,r^2_Z)}}\right]
		\nonumber \\
	&\approx & \frac{v^{\prime 2} m_{\phi}^3}{2 \pi v^4}
		\nonumber \\
	\Gamma (\phi^p\rightarrow t \bar{t}) &=& \frac{N_c}{4\pi }
	\frac{r^2_t m^3_\phi }{4f^2}
	\frac{\lambda^{\frac{1}{2}}(1,r^2_t ,r^2_t)}
	{\sqrt{4r_t^2 + \lambda (1,r^2_t ,r^2_t)}}\approx 
	\frac{N_c m^2_t m_\phi }{16\pi f^2} 
		\nonumber \\
	\Gamma (\phi^p\rightarrow b \bar{b}) &=& \frac{N_c}{4\pi }
	\frac{r^2_b m^3_\phi }{4f^2}
	\frac{\lambda^{\frac{1}{2}}(1,r^2_b ,r^2_b)}
	{\sqrt{4r_b^2 + \lambda (1,r^2_b ,r^2_b)}}\approx 
	\frac{N_c m^2_b m_\phi }{16\pi f^2} 
		\nonumber \\
	\Gamma (\phi^p\rightarrow T \bar{t} + t \bar{T}) &=& \frac{N_c}{4\pi }
	\frac{r^2_tm^3_\phi }{2f^2}
	\left[\frac{\lambda^{\frac{1}{2}}(1,r^2_T ,r^2_t) (1 - r_T^2 - r_t^2)}
	{\sqrt{4r_T^2 + \lambda (1,r^2_T ,r^2_t)} 
	+ \sqrt{4r_t^2 + \lambda (1,r^2_T ,r^2_t)}}\right]
	\left(\frac{\lambda_1}{\lambda_2}\right)^2
					\nonumber \\
	&\approx & \frac{N_c m^2_t m_\phi }{16 \pi f^2}
		\left(\frac{\lambda_1}{\lambda_2}\right)^2
		(1-r^2_T)^2.
\end{eqnarray}
In the $\phi^+$, $\phi^s$, $\phi^p$ couplings to quarks, we have 
neglected $v^{\prime}/v$ relative to $v/f$ and included the color
factor, $N_c = 3$.

\end{document}